\documentclass[superscriptaddress,showpacs,aps,prd]{revtex4}
\usepackage[paperwidth=210mm,paperheight=297mm,centering,hmargin=2cm,vmargin=4.0cm]{geometry}
\usepackage{graphicx}
\usepackage{multibib}
\usepackage{dcolumn}
\usepackage{amsmath}
\usepackage{wasysym}
\usepackage{amsfonts}
\usepackage{url}
\usepackage{hyperref}
\usepackage{color}
\usepackage{datatool}
\DTLsetseparator{;}
\def\ee{e^+e^-}
\def\ra{\rightarrow}
\def\pipi{\pi^+\pi^-}
\def\epp{\eta\pipi}
\def\SP{\ee\ra\epp}
\def\gg{\gamma\gamma}
\def\ro{\rho(770)}
\def\rop{\rho(1450)}
\def\ropp{\rho(1700)}

\DTLloaddb{cross-section-fit}{cross-section-fit.csv}
\DTLloaddb{products-rho1450}{products-rho1450.csv}
\DTLloaddb{products-rho1700}{products-rho1700.csv}
\DTLloaddb{cross-section-data}{cross-section-data.csv}
\DTLloaddb{cross-section-data-combined}{cross-section-data-combined.csv}
\begin{document}
\begin{filecontents*}{cross-section-fit.csv}
pars; units; m1; m2; m3; m4
g_{\rho(770)}; GeV$^{-1}$; 1.586; 1.586; 1.586; 1.586
g_{\rho(1450)}; GeV$^{-1}$; 0.40\pm0.03; 0.58\pm0.02; 0.36\pm0.02; 0.58\pm0.03
g_{\rho(1700)}; GeV$^{-1}$; ---; ---; (0.50\pm0.16)\times10^{-2}; (0.54\pm0.18)\times10^{-2}
M_{\rho(770)}; GeV; 0.775; 0.775; 0.775; 0.775
M_{\rho(1450)}; GeV; 1.532\pm0.010; 1.536\pm0.010; 1.502\pm0.011; 1.506\pm0.011
M_{\rho(1700)}; GeV; ---; ---; 1.835\pm0.011; 1.834\pm0.012
\Gamma_{\rho(770)}; GeV; 0.149; 0.149; 0.149; 0.149
\Gamma_{\rho(1450)}; GeV; 0.360\pm0.029; 0.367\pm0.030; 0.315\pm0.027; 0.321\pm0.027
\Gamma_{\rho(1700)}; GeV; ---; ---; (0.45\pm0.19)\times10^{-1}; (0.47\pm0.19)\times10^{-1}
\phi_{\rho(770)}; rad; 0; 0; 0; 0
\phi_{\rho(1450)}; rad; 2.25\pm0.20; 3.81\pm0.14; 1.73\pm0.20; 4.16\pm0.13
\phi_{\rho(1700)}; rad; ---; ---; 3.95\pm0.39; 0.81\pm0.52
\mathcal{B}(\rho(1450)\rightarrow\pi^{+}\pi^{-}); $\%$; 15; 15; 15; 15
\mathcal{B}(\rho(1700)\rightarrow\pi^{+}\pi^{-}); $\%$; ---; ---; 14; 14
\mathcal{B}(\rho(1450)\rightarrow\omega\pi^0); $\%$; 45; 45; 45; 45
\mathcal{B}(\rho(1700)\rightarrow\omega\pi^0); $\%$; ---; ---; 0; 0
\mathcal{B}(\rho(1450)\rightarrow4\pi); $\%$; 40; 40; 40; 40
\mathcal{B}(\rho(1700)\rightarrow4\pi); $\%$; ---; ---; 86; 86
\chi^2/ndf; ---; 98.8/79; 99.0/79; 72.1/75; 71.9/75
\end{filecontents*}

\begin{filecontents*}{products-rho1450.csv}
solution; wbproduct
Model 1, solution 1; $178\pm27\pm11$
Model 1, solution 2; $377\pm14\pm23$
Model 2, solution 1; $125\pm16\pm8$
Model 2, solution 2; $335\pm27\pm20$
\end{filecontents*}

\begin{filecontents*}{products-rho1700.csv}
solution; wbproduct
Model 2, solution 1; $1.21\pm0.47\pm0.07$
Model 2, solution 2; $1.35\pm0.53\pm0.08$
\end{filecontents*}

\begin{filecontents*}{cross-section-data.csv}
ecm; cs; nev; eff; lum; secm; scs; snev; seff; slum
1.1243; 0.05\pm0.12; 4\pm8; 0.337; 552    ;   1.7550; 2.46\pm0.36; 252\pm27; 0.249; 1030
1.1506; 0.00\pm0.17; 0\pm9; 0.327; 494  ;   1.7577; 2.45\pm0.39; 256\pm26; 0.272; 965 
1.1961; 0.00\pm0.17; 0\pm10; 0.308; 557 ;   1.7737; 2.28\pm0.45; 139\pm20; 0.271; 555 
1.2234; 0.24\pm0.21; 13\pm12; 0.333; 536  ;   1.7782; 1.64\pm0.33; 204\pm23; 0.244; 1119
1.2449; 0.31\pm0.30; 13\pm12; 0.314; 408  ;   1.7929; 1.88\pm0.51; 95\pm18; 0.272; 448  
1.2728; 0.16\pm0.19; 9\pm9; 0.341; 456	  ;   1.7980; 3.05\pm0.37; 289\pm26; 0.267; 990 
1.2771; 0.30\pm0.19; 23\pm12; 0.322; 720  ;   1.7988; 1.62\pm0.42; 174\pm22; 0.242; 938 
1.2822; 0.35\pm0.10; 110\pm26; 0.308; 3080;   1.8198; 1.70\pm0.28; 190\pm22; 0.237; 1120
1.2951; 0.47\pm0.25; 24\pm11; 0.335; 451  ;   1.8264; 2.16\pm0.47; 116\pm18; 0.272; 508 
1.2997; 0.66\pm0.21; 57\pm15; 0.308; 872  ;   1.8400; 1.41\pm0.30; 157\pm21; 0.257; 960 
1.3234; 0.58\pm0.26; 38\pm15; 0.353; 530  ;   1.8401; 1.05\pm0.42; 176\pm23; 0.231; 1336
1.3436; 0.86\pm0.28; 58\pm17; 0.352; 558  ;   1.8486; 0.88\pm0.42; 54\pm13; 0.262; 438  
1.3502; 1.02\pm0.28; 127\pm29; 0.304; 1217;   1.8600; 1.50\pm0.27; 214\pm26; 0.226; 1524
1.3565; 1.70\pm0.30; 152\pm23; 0.333; 843 ;   1.8712; 0.51\pm0.32; 58\pm15; 0.250; 664  
1.3735; 1.55\pm0.55; 33\pm10; 0.342; 181  ;   1.8718; 1.08\pm0.41; 108\pm21; 0.219; 1035
1.3940; 1.61\pm0.32; 101\pm17; 0.340; 527 ;   1.8743; 0.89\pm0.36; 92\pm19; 0.243; 851  
1.4013; 1.66\pm0.29; 155\pm22; 0.306; 871 ;   1.8748; 0.60\pm0.36; 93\pm20; 0.229; 1080 
1.4349; 3.81\pm0.33; 374\pm30; 0.321; 916 ;   1.8751; 0.54\pm0.29; 146\pm25; 0.216; 1884
1.4501; 3.68\pm0.36; 421\pm35; 0.297; 1107;   1.8766; 1.15\pm0.25; 269\pm33; 0.216; 2512
1.4715; 4.05\pm0.44; 236\pm22; 0.327; 509 ;   1.8778; 1.24\pm0.32; 233\pm33; 0.217; 2046
1.4997; 3.79\pm0.36; 436\pm35; 0.288; 1100;   1.8792; 0.94\pm0.28; 198\pm29; 0.215; 2007
1.5146; 4.47\pm0.42; 416\pm33; 0.315; 835 ;   1.8804; 0.73\pm0.25; 165\pm24; 0.217; 1888
1.5224; 4.45\pm0.50; 274\pm24; 0.322; 534 ;   1.8814; 1.45\pm0.30; 228\pm29; 0.216; 1856
1.5432; 4.73\pm0.59; 273\pm29; 0.313; 514 ;   1.8840; 0.68\pm0.32; 112\pm23; 0.217; 1315
1.5499; 4.83\pm0.50; 467\pm37; 0.279; 964 ;   1.8934; 1.34\pm0.39; 71\pm14; 0.248; 524  
1.5719; 4.10\pm0.51; 251\pm25; 0.311; 524 ;   1.9010; 1.17\pm0.33; 129\pm23; 0.218; 1158
1.5938; 3.47\pm0.54; 186\pm22; 0.309; 448 ;   1.9013; 1.89\pm0.60; 83\pm16; 0.246; 501  
1.5950; 4.10\pm0.51; 364\pm29; 0.295; 825 ;   1.9032; 1.13\pm0.38; 112\pm20; 0.244; 897 
1.6019; 3.47\pm0.39; 453\pm37; 0.273; 1234;   1.9212; 1.06\pm0.26; 134\pm22; 0.212; 1332
1.6229; 3.70\pm0.47; 218\pm22; 0.304; 513 ;   1.9248; 0.90\pm0.36; 58\pm13; 0.237; 566  
1.6430; 3.80\pm0.56; 195\pm23; 0.296; 459 ;   1.9270; 0.78\pm0.43; 57\pm15; 0.243; 592  
1.6503; 2.45\pm0.35; 373\pm33; 0.257; 1374;   1.9428; 0.79\pm0.20; 141\pm21; 0.211; 1754
1.6694; 2.70\pm0.45; 177\pm22; 0.292; 563 ;   1.9449; 0.77\pm0.34; 89\pm20; 0.240; 991  
1.6741; 2.31\pm0.40; 239\pm26; 0.280; 881 ;   1.9526; 1.31\pm0.47; 56\pm14; 0.240; 452  
1.6792; 3.04\pm0.57; 183\pm24; 0.251; 632 ;   1.9640; 0.94\pm0.29; 117\pm22; 0.210; 1304
1.6929; 2.79\pm0.47; 158\pm20; 0.291; 494 ;   1.9670; 0.73\pm0.35; 60\pm14; 0.233; 693  
1.7000; 1.81\pm0.36; 195\pm23; 0.253; 939 ;   1.9784; 0.95\pm0.37; 52\pm13; 0.237; 524  
1.7158; 2.65\pm0.40; 227\pm26; 0.273; 807 ;   1.9826; 0.89\pm0.30; 105\pm20; 0.211; 1229
1.7200; 2.20\pm0.42; 211\pm26; 0.253; 913 ;   1.9885; 1.06\pm0.37; 63\pm14; 0.232; 602  
1.7231; 2.80\pm0.48; 159\pm19; 0.286; 525 ;   2.0046; 1.03\pm0.46; 51\pm15; 0.239; 481  
1.7400; 1.68\pm0.33; 176\pm22; 0.250; 933 ;   2.0070; 0.82\pm0.19; 294\pm33; 0.204; 3732
1.7416; 2.94\pm0.51; 165\pm19; 0.287; 540 ; ; ; ; ;
\end{filecontents*}

\begin{filecontents*}{cross-section-data-combined.csv}
ecm; cs; secm; scs; tecm; tcs
1.1349\pm0.0129; 0.00\pm0.09; 1.5474\pm0.0032; 4.79\pm0.35; 1.8731\pm0.0016; 0.75\pm0.15
1.2076\pm0.0135; 0.08\pm0.13; 1.5881\pm0.0103; 3.88\pm0.26; 1.8770\pm0.0015; 0.95\pm0.13
1.2646\pm0.0127; 0.21\pm0.15; 1.6168\pm0.0162; 3.61\pm0.25; 1.8808\pm0.0005; 1.02\pm0.18
1.2810\pm0.0021; 0.33\pm0.08; 1.6626\pm0.0110; 2.53\pm0.20; 1.8875\pm0.0045; 0.95\pm0.24
1.2980\pm0.0022; 0.59\pm0.15; 1.6936\pm0.0081; 2.40\pm0.23; 1.9019\pm0.0010; 1.31\pm0.20
1.3326\pm0.0101; 0.71\pm0.18; 1.7195\pm0.0029; 2.52\pm0.22; 1.9226\pm0.0018; 1.00\pm0.20
1.3532\pm0.0032; 1.32\pm0.19; 1.7490\pm0.0080; 2.30\pm0.17; 1.9408\pm0.0061; 0.78\pm0.15
1.3887\pm0.0090; 1.61\pm0.27; 1.7885\pm0.0104; 2.11\pm0.15; 1.9631\pm0.0050; 0.96\pm0.18
1.4144\pm0.0164; 2.51\pm0.21; 1.8218\pm0.0030; 1.80\pm0.23; 1.9811\pm0.0020; 0.91\pm0.22
1.4729\pm0.0215; 3.87\pm0.21; 1.8401\pm0.0000; 1.31\pm0.19; 2.0040\pm0.0065; 0.90\pm0.13
1.5180\pm0.0039; 4.46\pm0.29; 1.8566\pm0.0052; 1.29\pm0.22; ;
\end{filecontents*}

\title{\large \bf \boldmath Measurement of the $\SP$ cross section
  with the CMD-3 detector at the VEPP-2000 collider}
\def\budkerinp{Budker Institute of Nuclear Physics, SB RAS, Novosibirsk,
  630090, Russia}
\def\nsu{Novosibirsk State University, Novosibirsk, 630090, Russia}
\def\lebedevpi{Lebedev Physical Institute RAS, Moscow, 119333, Russia}
\def\nstu{Novosibirsk State Technical University, Novosibirsk, 630092, Russia}
\def\hawaiiu{University of Hawaii, Honolulu, Hawaii 96822, USA}
\def\victoriau{University of Victoria, Victoria, BC, Canada V8W 3P6}
\def\infn{Istituto Nazionale di Fisica Nucleare, Sezione di Lecce, Lecce, Italy}

\author{S.S.~\surname{Gribanov}}
\email[E-mail:]{S.S.Gribanov@inp.nsk.su}
\affiliation{\budkerinp}
\affiliation{\nsu}

\author{A.S.~\surname{Popov}}
\affiliation{\budkerinp}
\affiliation{\nsu}

\author{R.R.~\surname{Akhmetshin}}
\affiliation{\budkerinp}
\affiliation{\nsu}

\author{A.N.~\surname{Amirkhanov}}
\affiliation{\budkerinp}
\affiliation{\nsu}

\author{A.V.~\surname{Anisenkov}}
\affiliation{\budkerinp}
\affiliation{\nsu}

\author{V.M.~\surname{Aulchenko}}
\affiliation{\budkerinp}
\affiliation{\nsu}

\author{V.Sh.~\surname{Banzarov}}
\affiliation{\budkerinp}

\author{N.S.~\surname{Bashtovoy}}
\affiliation{\budkerinp}

\author{D.E.~\surname{Berkaev}}
\affiliation{\budkerinp}
\affiliation{\nsu}

\author{A.E.~\surname{Bondar}}
\affiliation{\budkerinp}
\affiliation{\nsu}

\author{A.V.~\surname{Bragin}}
\affiliation{\budkerinp}

\author{S.I.~\surname{Eidelman}}
\affiliation{\budkerinp}
\affiliation{\nsu}
\affiliation{\lebedevpi}

\author{D.A.~\surname{Epifanov}}
\affiliation{\budkerinp}
\affiliation{\nsu}

\author{L.B.~\surname{Epshteyn}}
\affiliation{\budkerinp}
\affiliation{\nsu}
\affiliation{\nstu}

\author{A.L.~\surname{Erofeev}}
\affiliation{\budkerinp}
\affiliation{\nsu}

\author{G.V.~\surname{Fedotovich}}
\affiliation{\budkerinp}
\affiliation{\nsu}

\author{S.E.~\surname{Gayazov}}
\affiliation{\budkerinp}
\affiliation{\nsu}

\author{F.J.~\surname{Grancagnolo}}
\affiliation{\infn}

\author{A.A.~\surname{Grebenuk}}
\affiliation{\budkerinp}
\affiliation{\nsu}

\author{D.N.~\surname{Grigoriev}}
\affiliation{\budkerinp}
\affiliation{\nsu}
\affiliation{\nstu}

\author{F.V.~\surname{Ignatov}}
\affiliation{\budkerinp}
\affiliation{\nsu}

\author{V.L.~\surname{Ivanov}}
\affiliation{\budkerinp}
\affiliation{\nsu}

\author{S.V.~\surname{Karpov}}
\affiliation{\budkerinp}

\author{V.F.~\surname{Kazanin}}
\affiliation{\budkerinp}
\affiliation{\nsu}

\author{A.N.~\surname{Kirpotin}}
\affiliation{\budkerinp}

\author{I.A.~\surname{Koop}}
\affiliation{\budkerinp}
\affiliation{\nsu}

\author{A.A.~\surname{Korobov}}
\affiliation{\budkerinp}
\affiliation{\nsu}

\author{O.A.~\surname{Kovalenko}}
\affiliation{\budkerinp}

\author{A.N.~\surname{Kozyrev}}
\affiliation{\budkerinp}
\affiliation{\nstu}

\author{E.A.~\surname{Kozyrev}}
\affiliation{\budkerinp}
\affiliation{\nsu}

\author{P.P.~\surname{Krokovny}}
\affiliation{\budkerinp}
\affiliation{\nsu}

\author{A.E.~\surname{Kuzmenko}}
\affiliation{\budkerinp}
\affiliation{\nsu}

\author{A.S.~\surname{Kuzmin}}
\affiliation{\budkerinp}
\affiliation{\nsu}

\author{I.B.~\surname{Logashenko}}
\affiliation{\budkerinp}
\affiliation{\nsu}

\author{P.A.~\surname{Lukin}}
\affiliation{\budkerinp}
\affiliation{\nsu}

\author{K.Yu.~\surname{Mikhailov}}
\affiliation{\budkerinp}
\affiliation{\nsu}

\author{V.S.~\surname{Okhapkin}}
\affiliation{\budkerinp}

\author{A.V.~\surname{Otboev}}
\affiliation{\budkerinp}

\author{Yu.N.~\surname{Pestov}}
\affiliation{\budkerinp}

\author{G.P.~\surname{Razuvaev}}
\affiliation{\budkerinp}
\affiliation{\nsu}

\author{Yu.A.~\surname{Rogovsky}}
\affiliation{\budkerinp}

\author{A.A.~\surname{Ruban}}
\affiliation{\budkerinp}

\author{N.M.~\surname{Ryskulov}}
\affiliation{\budkerinp}

\author{A.E.~\surname{Ryzhenenkov}}
\affiliation{\budkerinp}
\affiliation{\nsu}

\author{A.V.~\surname{Semenov}}
\affiliation{\budkerinp}
\affiliation{\nsu}

\author{A.I.~\surname{Senchenko}}
\affiliation{\budkerinp}

\author{Yu.M.~\surname{Shatunov}}
\affiliation{\budkerinp}

\author{P.Yu.~\surname{Shatunov}}
\affiliation{\budkerinp}

\author{V.E.~\surname{Shebalin}}
\affiliation{\budkerinp}
\affiliation{\nsu}
\affiliation{\hawaiiu}

\author{D.N.~\surname{Shemyakin}}
\affiliation{\budkerinp}
\affiliation{\nsu}

\author{B.A.~\surname{Shwartz}}
\affiliation{\budkerinp}
\affiliation{\nsu}

\author{D.B.~\surname{Shwartz}}
\affiliation{\budkerinp}
\affiliation{\nsu}

\author{A.L.~\surname{Sibidanov}}
\affiliation{\budkerinp}
\affiliation{\victoriau}

\author{E.P.~\surname{Solodov}}
\affiliation{\budkerinp}
\affiliation{\nsu}

\author{V.M.~\surname{Titov}}
\affiliation{\budkerinp}

\author{A.A.~\surname{Talyshev}}
\affiliation{\budkerinp}
\affiliation{\nsu}

\author{S.S.~\surname{Tolmachev}}
\affiliation{\budkerinp}

\author{A.I.~\surname{Vorobiov}}
\affiliation{\budkerinp}

\author{I.M.~\surname{Zemlyansky}}
\affiliation{\budkerinp}

\author{Yu.V.~\surname{Yudin}}
\affiliation{\budkerinp}
\affiliation{\nsu}

\date{\today}
 
\begin{abstract}
  The cross section of the process $\SP$ is measured using the
  data collected with the CMD-$3$ detector at the VEPP-$2000$ collider
  in the center-of-mass energy range from $1.1$ to $2.0$ GeV. The 
  decay mode $\eta\ra\gg$ is used for $\eta$ meson reconstruction in
  the data sample corresponding to an integrated luminosity of $78.3$ pb$^{-1}$.
  The energy dependence of the $\SP$ cross section is fitted within the 
  framework of vector meson dominance in order to extract the 
  $\Gamma(\rop\ra\ee)\mathcal{B}(\rop\ra\epp)$ and
  the $\Gamma(\ropp\ra\ee)\mathcal{B}(\ropp\ra\epp)$ products.
  Based on conservation of vector current, the analyzed data are used to 
  test the relationship between the $\SP$ cross section and the spectral 
  function in $\tau^-\ra\eta\pi^-\pi^0\nu_\tau$ decay. The $\SP$ cross section 
  obtained with the CMD-$3$ detector is in good agreement with the previous
  measurements.
\end{abstract}

\maketitle

\section{Introduction}
We report on a study of the process $\SP$ with the CMD-$3$
  detector at the VEPP-$2000$ $e^+e^-$ collider, where $\eta$ mesons
  are reconstructed using  the decay mode $\eta\ra\gg$. In the previous
experiments it has been shown
  that this isovector final state is mainly produced through
  the $\eta\ro$ intermediate mechanism~\cite{snd-2015, snd-2018}. As a part 
of the total hadronic cross section, the cross section of the process $\SP$ 
is interesting for the calculations of the hadronic contribution to the muon 
anomalous magnetic moment~\cite{davier-2017,jegerlenher-2018,keshavarzi-2018}.
The $\SP$ cross section data can be also used to study the properties of
the $\rop$ and $\ropp$ resonances, as well as to obtain the hadronic
spectral function for the $\tau^-\ra\eta\pi^-\pi^0\nu_\tau$ decay
and thus test conservation of vector current~\cite{cvc-2001}.

The process $\SP$ was studied earlier in several
experiments~\cite{nd-1986,dm2-1988,cmd2-2000,babar-2008,snd-2010,snd-2015,snd-2018,babar-2018-isr,babar-2018-3pi}.
The most precise measurements of its cross section have been performed
at the PEP-II B-factory by the BaBar Collaboration~\cite{babar-2018-isr}.

\section{Experiment}
\begin{figure}[t]
  \centering \includegraphics[width=0.8\linewidth]{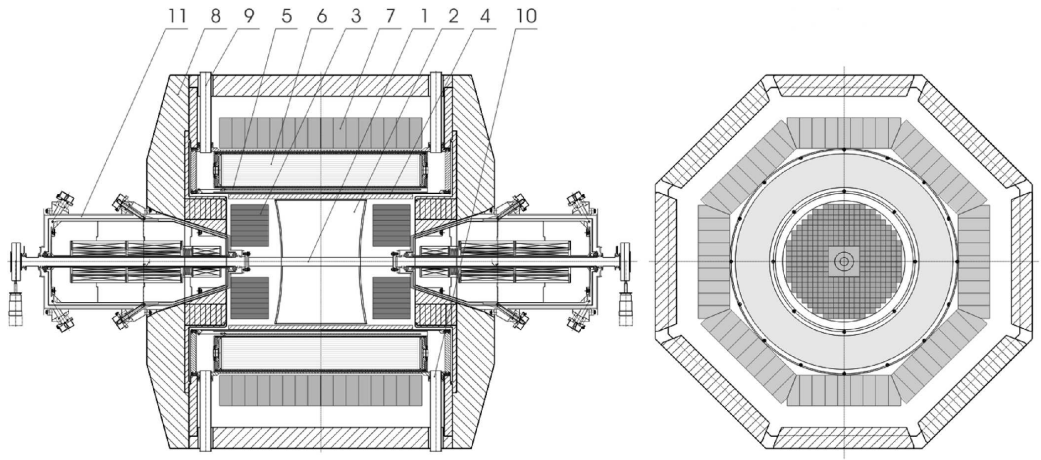}
\caption{The schematic view of the CMD-3 detector. (1) beam pipe,
  (2) drift chamber, (3) BGO endcap calorimeter, (4) Z-chamber, (5)
  superconducting solenoid, (6) liquid xenon calorimeter, (7) CsI
  barrel calorimeter, (8) iron yoke, (9) liquid He supply, (10) vacuum
  pumpdown, (11) VEPP-2000 superconducting magnetic lenses.
\label{fig:cmd3}}
\end{figure}
The data sample has been collected with the CMD-3 detector at the VEPP-$2000$
$e^+e^-$ collider~\cite{vepp-danilov-1996, vepp-koop-2008, vepp-compton-2013, vepp-compton-2015, vepp-shatunov-2016, vepp-shwartz-2016} in $2011$, $2012$ and $2017$ experimental runs.
In order to reach the design luminosity in the single-bunch mode, the collider 
is operated using the round beam technique~\cite{vepp-berkaev-2012}
in the center-of-mass (c.m.) energy range from $0.32$ to $2.0$ GeV. The beam 
energy was measured using a VEPP-2000 magnetic field in the $2011$ and $2012$
experimental runs~\cite{vepp-danilov-1996, vepp-koop-2008, vepp-shatunov-2016, vepp-shwartz-2016},
and with the backscattering-laser-light system in the $2017$
one~\cite{vepp-compton-2013, vepp-compton-2015}.
The accuracy of the beam energy measurements is about $3$ MeV in 2011
and about $1$ MeV in 2012,
while in 2017 it is better than $0.1$ MeV.
 
The general-purpose cryogenic magnetic detector CMD-3 has been
described in detail elsewhere~\cite{sndcmd3}. The schematic
view of the CMD-$3$ detector is shown in Fig.\ \ref{fig:cmd3}.
The tracking system of the CMD-$3$ detector consists of a double-layer
multiwire proportional Z-chamber~\cite{zc} and a cylindrical drift
chamber~\cite{dc} with hexagonal cells, which volume is filled with the
argon-isobutane gas mixture. Magnetic field of $1.3$ T inside the tracking 
system is provided by the superconducting solenoid, which surrounds the 
drift and Z-chambers. The barrel electromagnetic
calorimeter is situated outside the superconducting solenoid and consists of
two parts. The first part is the liquid xenon calorimeter (a thickness 
is $5.4X_0$, where $X_0$ is a radiation length), which allows photon 
coordinates to be measured with the accuracy of $1$--$2$ mm~\cite{lxe}.  
The second part is the calorimeter composed of CsI(Tl) and CsI(Na) 
crystals (a thickness of $8.1X_0$).
This calorimeter consists of $8$ octants and contains $1152$ counters. The
endcap calorimeter~\cite{cal} consists of two identical endcaps, each
containing $340$ BGO crystals with a thickness of $13.4X_0$.

\section{Simulation\label{sec:mc-simulation}}
The Monte Carlo (MC) simulation of the process $e^+e^-\ra\eta\pipi$
has been performed separately at each $e^+e^-$ energy corresponding to
the collected experimental data. It takes into account the $\eta\ro$ 
intermediate
state with the following matrix element:
\begin{align}
  M_{fi}\propto\frac{1}{D(Q_{\pipi})}
  \varepsilon_{\alpha\beta\gamma\delta} J^{\alpha}P^{\beta}_{\pi^+}
  P^{\gamma}_{\pi^-}P^{\delta}_{\eta},
  \label{eq:simulation-matrix-element}
\end{align}
where $J$ is a lepton current, $P_{\pi^+}$, $P_{\pi^-}$, $P_{\eta}$
are four-momenta of $\pi^+$, $\pi^-$ and $\eta$, respectively.
$D(Q_{\pipi}) = Q^2_{\pipi} - m^2_{\ro} + i
\sqrt{Q^2_{\pipi}}\Gamma_{\ro}$ is the inverse propagator of the
$\ro$, $m_{\ro}$ and $\Gamma_{\ro}$ are the mass and the width of the $\ro$,
respectively, and $Q_{\pipi} = P_{\pi^+} +P_{\pi^-}$ is its four-momentum.
To take into account the initial-state
radiation according to works~\cite{kuraev-fadin,actis}, the simulation is
done in two iterations. In the first iteration, the cross section
of the process $e^+e^-\ra\eta\pipi$ measured with BaBar is used to
simulate ISR photons, while in the second one the cross section
measured with the CMD-3 obtained in the first iteration is employed for this 
purpose. For a simulation of various multihadronic
backgrounds the MHG$2000$ generator specially developed for experiments at
CMD-$3$ has been used~\cite{cmd3-multihadron}. The interactions of the
generated particles with the detector and its response are implemented
using the Geant$4$ toolkit~\cite{geant4}.

\section{Event selection\label{sec:event-selection}}

To select $\SP$ event candidates, the following criteria are used.
To begin with, events are selected with two oppositely charged particles 
originating from the beam interaction region. In addition, it is required 
that the selected events contain at least two photons with energies greater 
than $50$ MeV to suppress background processes with low-energy
photons. Also excluded are photons, which pass through the BGO crystals 
closest to the beam axis.
For each selected event all photon pairs are considered and a kinematic fit 
is performed within the $\ee\ra\pipi\gg$ hypothesis using the constraints of 
energy-momentum conservation and requiring all
particles to originate from a common vertex. The photons from the pair 
corresponding to the smallest chi-square of the kinematic fit, 
$\chi^2_{\pipi\gg}$, are considered as candidates for the photons from the
$\eta\rightarrow\gamma\gamma$ decay. Only events with the fit quality 
$\chi^2_{\pipi\gg}<30$ are used
to obtain two-photon invariant mass spectra, discussed in 
Sec.~\ref{sec:yield}. The same condition was imposed
on the chi-square of the kinematic fit to obtain the distributions discussed 
in Sec.~\ref{sec:int_struct}.

\begin{figure*}
  \centering \includegraphics[width=0.8\linewidth]{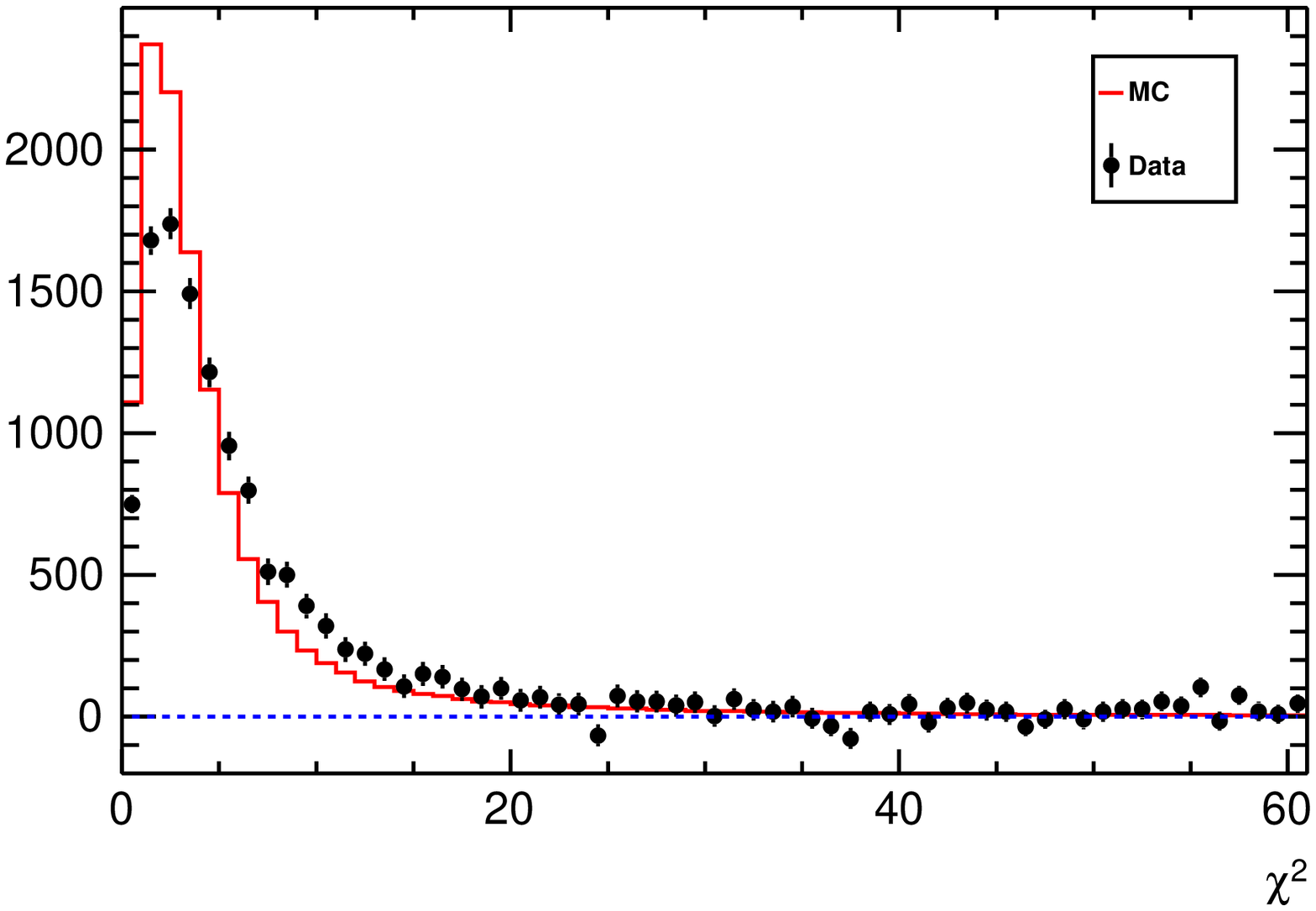}
  \caption{$\chi^2$ of the kinematic fit under the $\ee\ra\pipi\gg$
    hypothesis (points with error bars) and simulated $\SP$ events
    (histogram) from the energy range $\sqrt{s} = 1.3$--$1.8$
    GeV. \label{fig:kf_chi2}}
\end{figure*}

The $\chi^2_{\pipi\gg}$ distribution obtained using the whole $\SP$ data 
sample is shown in Fig.\ \ref{fig:kf_chi2}. The corresponding
$\chi^2$ distribution for simulated $\SP$ events is also shown. 
The contributions to
the $\chi^2_{\pipi\gg}$ distribution for simulated $\SP$ events at
each c.m.\ energy are proportional to $\sigma(\SP){L_{\rm int}}$,
where $\sigma(\SP)$ is the cross section of the process $\SP$
and $L_{\rm int}$ is the integrated luminosity. Efficiency corrections
discussed in Sec.~\ref{sec:efficiency} are also taken into account to obtain
the $\chi^2_{\pipi\gg}$ distribution for the MC data sample. In addition, 
these efficiency corrections are taken into account to obtain other MC 
distributions given in this paper.
The $\chi^2_{\pipi\gg}$ distributions have been obtained using all selection
criteria above except that on $\chi^2$ of the kinematic fit. The
remaining background (Sec.\ \ref{sec:yield}) is subtracted using
sidebands in two-photon invariant mass spectra
(Sec.\ \ref{sec:int_struct}). The histogram for simulated $\SP$ events
is normalized according to the ratio of the number of simulated and
experimental data events at $\chi^2<30$.  There is some disagreement between
$\chi^2$ distributions for the experimental data and simulated $\SP$ events. 
To address this disagreement, a corresponding correction to
the detection efficiency is applied, which is discussed in
Sec.\ \ref{sec:efficiency}.

\section{$\boldsymbol{\eta\pipi}$ event yield and background subtraction\label{sec:yield}}
To determine the $\epp$ yield the two-photon
invariant mass spectrum at each $e^+e^-$ energy in the experimental data is
fit with a sum of signal and background
distributions. The shape of the background distribution has been described
using a first-order polynomial. The shape of the signal
distribution has been fixed from the $\SP$ MC simulation using a
function, which is a linear combination of three Gaussian
distributions.

To take into account a difference in the two-photon mass resolution
and the $\eta$-meson peak position between the data and MC, two additional 
parameters, $\Delta{m}$ and
$\Delta\sigma^2$, are introduced. Here $\Delta{m}$ is the mass shift of the 
signal distribution as a whole and $\Delta\sigma^2$ is the square of the
two-photon mass resolution correction, which is added to the variance,
$\sigma^2$, of each Gaussian distribution from the signal function.

\begin{figure*}
  \centering \includegraphics[width=0.8\linewidth]{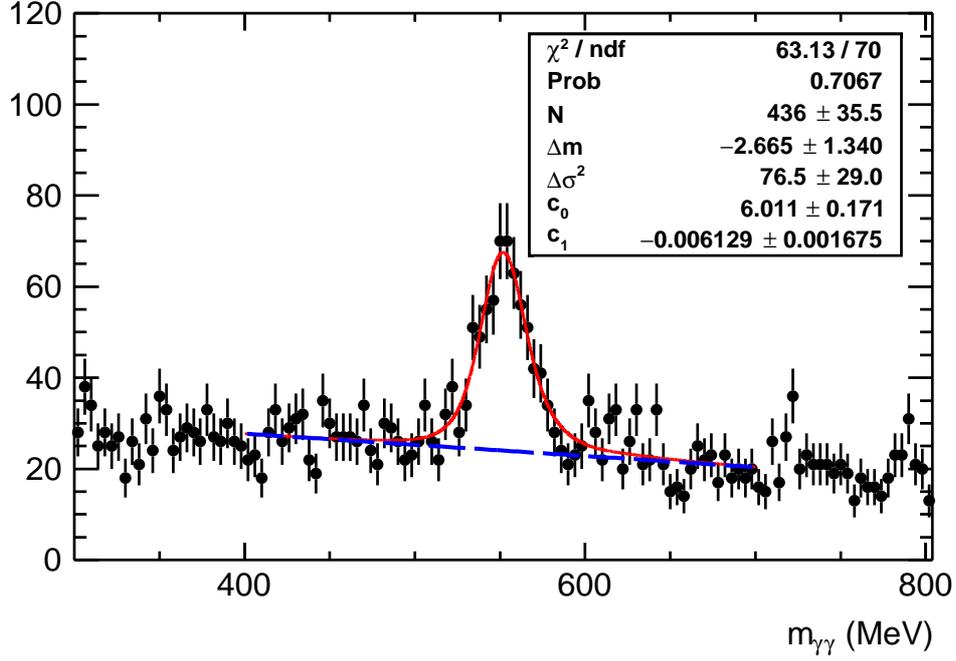}
  \caption{Two-photon invariant mass spectrum for the experimental
    data events (points with error bars) at $\sqrt{s} = 1.5$ GeV fitted 
with the function
    (solid curve), which contains the signal and background (dashed
    curve) contributions.\label{fig:mgg_data}}
\end{figure*}

The free parameters of the fit to the two-photon invariant mass spectrum are
the number of signal events, the mass shift of the signal, the square
of the two-photon mass resolution correction and background
distribution parameters. The total number of the fitted $\SP$ events
is $13426\pm206$. An example of the two-photon invariant mass spectrum
for $\SP$ event candidates at $\sqrt{s} = 1.5$ GeV is shown in
Fig.\ \ref{fig:mgg_data}.  The $\SP$ event yields for different
c.m.\ energy points are listed in Table~\ref{tab-result}. No excess of 
signal events over background is observed at c.m.\ energies
below $1.24$ GeV.

The main background source for the studied process is that with four 
final pions, $\ee\ra\pipi\pi^0\pi^0$.  Events of this process are
partially suppressed by selection criteria and do not have a peak at
the $\eta$-meson mass.  The sources of the peaking background, the
processes $\ee\ra\eta K^+K^-$ and $\ee\ra\epp\pi^0$, are strongly
suppressed by selection criteria. The contributions of these processes
have been estimated using MC simulation and corresponding cross
sections measured in Ref.\ \cite{babar-2008} and
Ref.\ \cite{cmd3-pi0epp}, respectively.  The contribution of each
process is found to be less than $0.1\%$ and neglected.

\section{Internal structure of the $\boldsymbol{\epp}$\label{sec:int_struct}}
\begin{figure}
  \centering \includegraphics[width=0.8\linewidth]{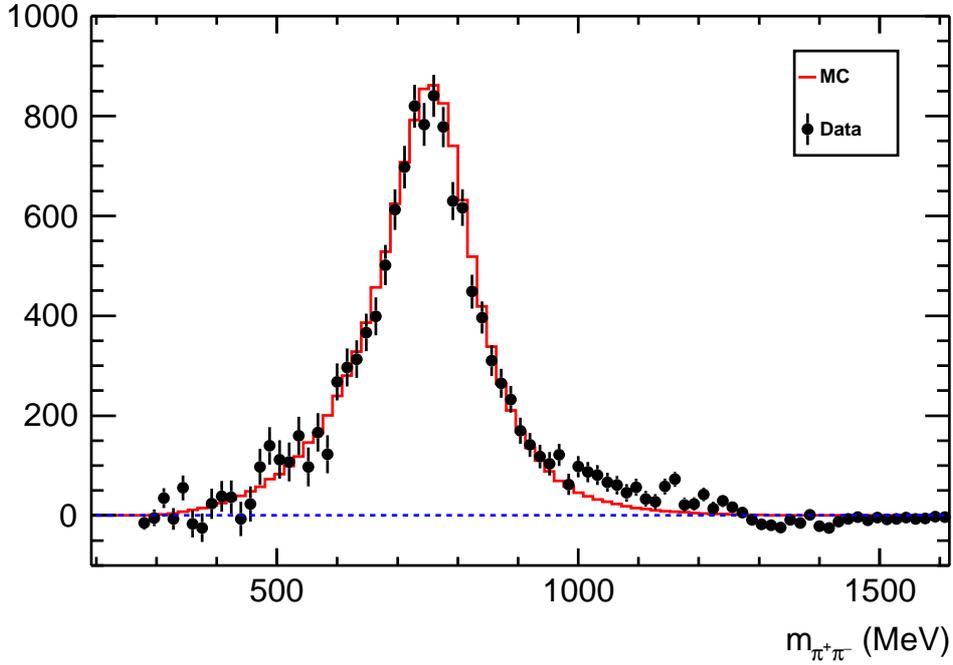}
  \caption{$\pipi$ invariant mass spectra for the experimental data 
    (points with error
    bars) and simulated $\SP$ events (histogram) from the energy range
    $\sqrt{s} = 1.3$--$1.8$ GeV. The simulation uses a model of the
    $\eta\ro$ intermediate state.\label{fig:mpipi}}
\end{figure}

The $\pipi$ invariant mass spectra for the whole $\SP$ data sample and 
simulated $\SP$ events
have been obtained as a difference between the $\pipi$ mass spectrum with 
$500\;{\rm MeV}<m_{\gg}<600\;{\rm MeV}$ and the spectrum for events
from sidebands ($400\;{\rm MeV}<m_{\gg}<470\;{\rm MeV}$
and $630\;{\rm MeV}<m_{\gg}<700\;{\rm MeV}$) divided by a
normalization factor of $1.4$. The $\pipi$ invariant mass spectra for
the whole $\SP$ data and simulated $\SP$ events are shown in
Fig.\ \ref{fig:mpipi}.  Points with error bars correspond to the
$\pipi$ invariant mass distribution for the whole $\SP$ data. The solid
histogram corresponds to the $\pipi$ invariant mass spectrum for
simulated $\SP$ events. The $\ro$ signal is seen in both
distributions. The contributions to the $\pipi$ invariant mass
spectrum for simulated $\SP$ events at each c.m.\ energy are
proportional to $\sigma(\SP){L_{\rm int}}$.
Since $\pipi$ spectra from data and simulation are very similar, we
can make a conclusion that the $\eta\ro$ intermediate mechanism
assumed in simulation gives indeed the dominant contribution to the
internal structure of the $\epp$ final state.

\begin{figure}
    \centering
    \includegraphics[width=0.8\linewidth]{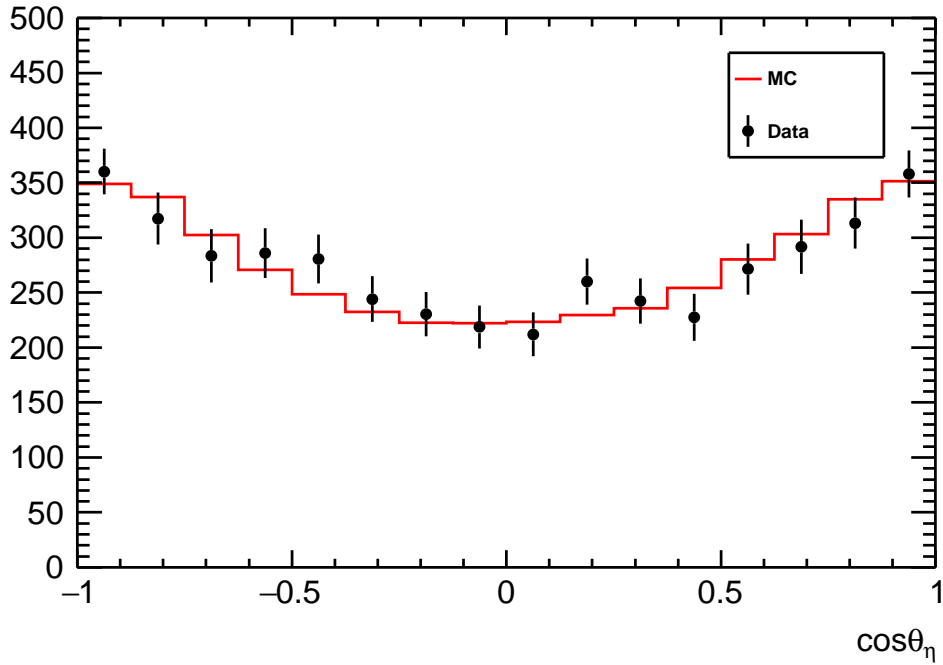}
    \caption{$\cos\theta_{\eta}$ distribution for the experimental data 
      (points with
      error bars) and simulated $\SP$ events (histogram) in the
      energy range $\sqrt{s} = 1.3$--$1.8$ GeV.  Simulation uses a
      model of the $\eta\ro$ intermediate state.\label{fig:cos_etath}}
\end{figure}

The distributions of the $\eta$-meson polar angle, $\theta_{\eta}$, for 
$\SP$ data and simulated $\SP$ events have been obtained in the same way as the
$\pipi$ invariant mass distributions and are shown in
Fig.\ \ref{fig:cos_etath}.  This distribution is expected to be proportional 
to $1 + \cos^2\theta_{\eta}$ in a model of the $\eta\rho$ intermediate state,
but the obtained one has a different shape because of the detector response.
The $\theta_{\eta}$ distributions for the whole $\SP$ data and
for simulated $\SP$ events are shown by points with error bars and by
a solid histogram, respectively.

\section{Detection efficiency \label{sec:efficiency}}
The detection efficiency for the process $\SP$ has been found from
corresponding MC simulation using the following formula:
\begin{align}
 \varepsilon_{\rm MC} = \frac{N^{\prime}_{\rm MC}}{N_{\rm MC}},
\end{align} 
where $N_{\rm MC}$ is the initial number of $\SP$ events generated
with the MC simulation and $N^{\prime}_{\rm MC}$ is the number of $\SP$ events 
extracted from the fit to the two-photon invariant mass spectrum.
\begin{figure}
  \centering \includegraphics[width=0.8\linewidth]{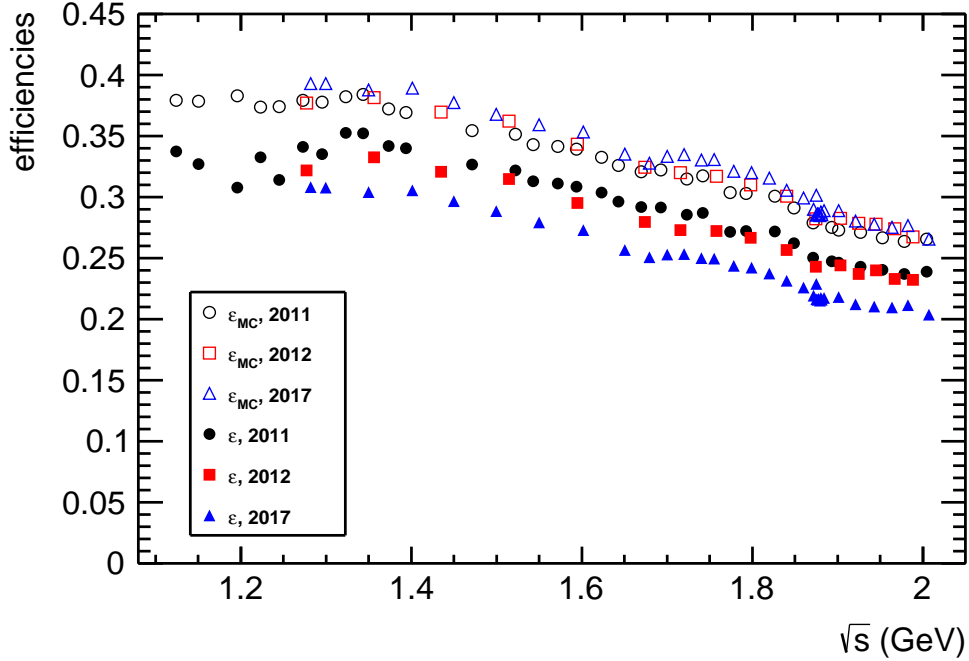}
  \caption{The $\SP$ detection efficiencies for $2011$, $2012$ and
    $2017$ data samples. The $\SP$ Monte Carlo detection efficiencies
    are indicated with empty markers while those with corrections 
    with filled markers.
\label{fig:efficiency}}
\end{figure}

To take into account the difference between the experimental data and the 
simulation, a set of corrections is applied to the detection efficiency 
found from the $\SP$ MC simulation. The corrected detection efficiency 
has been calculated using the following formula:
\begin{align}
  \label{eq:total-efficiency}
  \varepsilon = \varepsilon_{\rm MC} (1 + \delta_{\rm trigg})(1 +
  \delta_{\chi^2})(1 + \delta_{\pi})(1 + \delta_{\gamma}),
\end{align}
where $\delta_{\rm trigg}$ is the correction for trigger,  $\delta_{\pi}$ 
is the correction for charged pions, $\delta_{\gamma}$ is the correction
for photons and $\delta_{\chi^2}$ is the correction, which takes into
account a difference between the $\chi^2$ value of the kinematic fit 
distributions in data and the $\SP$ MC simulation. The energy dependence 
of the MC and corrected detection efficiencies is shown in
Fig.\ \ref{fig:efficiency}.

Events are recorded when a signal from at least one of the two
independent trigger systems is detected. One of these systems, the charged
trigger, uses information from the tracking system only, while the second one,
referred to as the neutral trigger,
is based on information from the electromagnetic calorimeter only.
The efficiencies of charged, $\varepsilon_{\rm CT}$, and neutral,
$\varepsilon_{\rm NT}$, triggers can be calculated using the following
relation:
\begin{align}
  \varepsilon_{\rm CT} = \frac{\rm N_{\rm CN}}{N_{\rm CN} + N_{\rm
      N}},\\ \varepsilon_{\rm NT} = \frac{N_{\rm CN}}{N_{\rm CN} +
    N_{\rm C}},\nonumber
\end{align}
where $N_{\rm CN}$ is the number of events with the simultaneous
signals from the charged and neutral triggers, $N_{\rm C}$ is the number
of events with signals from the charged trigger only and $N_{\rm N}$ is
the number of events with signals from the neutral one only. The
trigger efficiency correction, $\delta_{\rm trigg}$, can be calculated
using the trigger efficiencies in the following way:
\begin{align}
  \delta_{\rm trigg} = - (1 - \varepsilon_{\rm CT}) (1 -
  \varepsilon_{\rm NT}).
  \label{eq:trigger-correction}
\end{align}
The typical values of the trigger efficiency correction at
$\sqrt{s}>1.35$ GeV are about $(-0.9\pm0.1)\%$ and $(-1.0\pm0.1)\%$
for the $2011$ and $2012$ data samples, respectively, while 
at $\sqrt{s}\leq1.35$ GeV they are
$(-3.8\pm0.9)\%$ and $(-3.4\pm0.7)\%$. The typical value of the trigger 
efficiency correction for the $2017$ data sample is $(-0.58\pm0.06)\permil$.

The correction, which takes into account a difference between the
$\chi^2$ value of the kinematic fit distributions for the experimental data and
the simulated $\SP$ events, has been calculated using the numbers of $\SP$ 
events in two statistically independent regions 
$\chi^2_{\pi^+\pi^-\gamma\gamma}<30$ and
$30\leq\chi^2_{\pi^+\pi^-\gamma\gamma}<50$. An additional selection
criterion $N_{\gamma}=2$ is also used, where $N_{\gamma}$ is the number
of photons that are candidates for the $\eta$ decay photons. All other
selection criteria are the same as described in Sec.~\ref{sec:event-selection}.
The correction is given by the following equation:
\begin{align}
  \delta_{\chi^2} = 1 - (1 + \Delta N / N)_{\rm data} / (1 + \Delta N
  / N)_{\rm MC},
  \label{eq:chi2-correction}
\end{align}
where $N$ is the number of events in the region
$\chi^2_{\pi^+\pi^-\gamma\gamma}<30$ and $\Delta N$ is the number of
events in the region $30<\chi^2_{\pi^+\pi^-\gamma\gamma}<50$. The numbers
$N$ and $\Delta N$ are found using the $m_{\gamma\gamma}$ spectrum fitting
procedure described in the Sec.~\ref{sec:yield}.
The corresponding detection efficiency corrections, $\delta_{\chi^2}$'s,
are $(-1.6\pm0.7)\%$, $(-3.4\pm1.1)\%$ and $(-3.3\pm0.8)\%$ for the
$2011$, $2012$ and $2017$ data samples, respectively.

The charged-pion detection efficiency correction, $\delta_{\pi}$, has
been calculated using the following relation:
\begin{align}
  1 + \delta_{\pi} = \sum\Big(N^{\rm data}_{\pi^+}(\theta_{\pi^+}) /
  N^{\rm MC}_{\pi^+}(\theta_{\pi^+})\Big) \Big(N^{\rm
    data}_{\pi^-}(\theta_{\pi^-}) / N^{\rm
    MC}_{\pi^-}(\theta_{\pi^-})\Big)/N_{\rm MC},
  \label{eq:track-correction}
\end{align}
where the sum is taken over events from the $\SP$ MC simulation,
$N_{\rm MC}$ is the number of simulated $\SP$ events,
$N_{\pi^{\pm}}(\theta_{\pi^{\pm}})$ is the number of $\pi^{\pm}$
tracks with the polar angle equal to $\theta_{\pi^{\pm}}$ in the case,
when the second track hits the barrel part of the electromagnetic
calorimeter. The superscripts $\rm data$ and $\rm MC$ correspond to the
experimental and simulated $\SP$ events. 
The $N^{\rm MC}_{\pi^{\pm}}(\theta_{\pi^{\pm}})$
distribution is normalized to the number of events in the 
$ N^{\rm data}_{\pi^{\pm}}(\theta_{\pi^{\pm}})$
distribution inside the polar angle region corresponding to the barrel part 
of the calorimeter.
Since the reconstruction efficiency for the second track is close to 
$99\%$~\cite{kskl-2016}, 
the ratio of the number of events $N^{\rm data}_{\pi^{\pm}}(\theta_{\pi^{\pm}})/N^{\rm
  MC}_{\pi^{\pm}}(\theta_{\pi^{\pm}})$ is close to the ratio
$\varepsilon^{\rm
  data}_{\pi^{\pm}}(\theta_{\pi^{\pm}})/\varepsilon^{\rm
  MC}_{\pi^{\pm}}(\theta_{\pi^{\pm}})$, where
$\varepsilon_{\pi^{\pm}}$ is the $\pi^{\pm}$ reconstruction
efficiency.  The typical values of this correction are about
$(-6\pm6)\%$, $(-9\pm6)\%$ and $(-20\pm4)\%$ for the $2011$, $2012$
and $2017$ data samples, respectively.

The photon detection efficiency correction, $\delta_{\gamma}$, has
been calculated using the ratio of the reconstruction efficiencies of
photons in data and simulation, $\varepsilon^{\rm
  data}_{\gamma}(\theta)/\varepsilon^{\rm MC}_{\gamma}(\theta)$:
\begin{align}
  1 + \delta_{\gamma} = \sum\Big(\varepsilon^{\rm
    data}_{\gamma}(\theta_{\gamma_1}) / \varepsilon^{\rm
    MC}_{\gamma}(\theta_{\gamma_1})\Big) \Big(\varepsilon^{\rm
    data}_{\gamma}(\theta_{\gamma_2}) / \varepsilon^{\rm
    MC}_{\gamma}(\theta_{\gamma_2})\Big)/N_{\rm MC},
  \label{eq:photon-correction}
\end{align}
where the sum is taken over events from the $\SP$ MC simulation.  The
typical value of this correction is about $(-0.8\pm0.2)\%$ for the $2011$,
$2012$ and $2017$ data samples. The ratio of the photon
reconstruction efficiencies for the experimental data and the simulated events,
$\varepsilon^{\rm data}_{\gamma}(\theta)/\varepsilon^{\rm MC}_{\gamma}(\theta)$, 
has been found using events of the
process $\ee\ra\pipi\pi^0$. Photon reconstruction efficiencies for
both  data and simulated $\ee\ra\pipi\pi^0$ events have been
calculated as the following ratio:
\begin{align}
  \varepsilon_{\gamma}(\theta) =
  \frac{N^{\gamma}_2}{N^{\gamma}_1 +
    N^{\gamma}_2},
\end{align}
where $N^{\gamma}_1$ is the number of events where only one photon has
been detected in the barrel part of the calorimeter and
$N^{\gamma}_2$ is the number of events with two photons detected:
one of them in the barrel part of the calorimeter and the second one in the 
polar angle $\theta$.

\section{Results and discussion}
\begin{table}
  \caption{The c.m.\ energy ($\sqrt{s}$), the $\SP$ Born cross section
    ($\sigma_{\rm B}$), number of selected signal events ($N$),
    detection efficiency ($\varepsilon$), integrated luminosity
    ($L_{\rm int}$).}
  \label{tab-result}
  \begin{ruledtabular}
    \begin{tabular}[t]{ccccc|ccccc}
      $\sqrt{s}$, GeV &$\sigma_{\rm B}$, nb &$N$ &$\varepsilon$
      &$L_{\rm int}$, nb$^{-1}$& $\sqrt{s}$, GeV &$\sigma_{\rm B}$, nb
      &$N$ &$\varepsilon$ &$L_{\rm int}$, nb$^{-1}$\\ [0.5ex] \hline
      \DTLforeach{cross-section-data}{\ecm=ecm, \csec=cs, \nev=nev,
        \eff=eff, \lum=lum, \secm=secm, \scsec=scs, \snev=snev,
        \seff=seff, \slum=slum} { $\ecm$ & $\csec$ & $\nev$ & $\eff$ &
        $\lum$ & $\secm$ & $\scsec$ & $\snev$ & $\seff$ & $\slum$
        \DTLiflastrow{\\ [-2.5ex]}{\\} }
    \end{tabular}
  \end{ruledtabular}
\end{table}
\begin{table}
\caption{The $\SP$ Born cross section ($\sigma_{\rm B}$) at combined
  c.m.\ energies ($\sqrt{s}$).}
\label{tab-cs-combined}
  \begin{ruledtabular}
    \begin{tabular}[t]{cc|cc|cc}
      $\sqrt{s}$, GeV &$\sigma_{\rm B}$, nb & $\sqrt{s}$, GeV &
      $\sigma_{\rm B}$, nb & $\sqrt{s}$, GeV & $\sigma_{\rm B}$,
      nb\\ [0.5ex] \hline
      \DTLforeach{cross-section-data-combined}{\ecm=ecm,
        \csec=cs,\secm=secm, \scsec=scs, \tecm=tecm, \tcsec=tcs} {
        $\ecm$ & $\csec$ & $\secm$ & $\scsec$&$\tecm$ & $\tcsec$
        \DTLiflastrow{\\ [-2.5ex]}{\\} }
    \end{tabular}
  \end{ruledtabular}
\end{table}
The visible cross section at each c.m.\ energy has been calculated
using the following formula:
\begin{align}
  \sigma_{\rm vis} = \frac{N}{L_{\rm int}},
  \label{eq:visible-cross-section}
\end{align}
where $N$ is the $\SP$ yield and $L_{\rm int}$ is an integrated
luminosity.  The integrated luminosity at each c.m.\ energy has been
measured using the $\ee\ra\ee$ events~\cite{lum}. The visible and Born
cross sections are related by the following
equation~\cite{kuraev-fadin}:
\begin{align}
  \sigma_{\rm vis}(s) = \int\limits^{x_0}_0 dx\; \sigma_{\rm
    B}(s(1-x))\varepsilon(x, s)F(x,
  s), \label{eq:fadin_kuraev_formula_0}\\ x_0 = 1 -
  (2m_{\pi}+m_{\eta})^2/s,\nonumber
\end{align}
where $\sigma_{\rm vis}$ and $\sigma_{\rm B}$ are the visible and Born
cross sections, respectively. Here $F(x,s)$ is the initial-state
radiation (ISR) kernel function, $\varepsilon(x, s)$ is the detection
efficiency, which depends on the fraction of energy carried away by an ISR
photon, $m_{\pi}$ and $m_{\eta}$ are masses of the $\pi$ meson and
$\eta$ meson, respectively. The detection efficiency for events of the
$\SP$ MC simulation at each c.m.\ energy can be written in the
following form:
\begin{align}
  \varepsilon(s) = \frac{\int\limits^{x_0}_0 dx\; \sigma_{\rm
      B}(s(1-x))\varepsilon(x, s)F(x, s)}{\int\limits^{x_0}_0 dx\;
    \sigma_{\rm B}(s(1-x))F(x, s)}. \label{eq:eps_fadin_kuraev}
\end{align}
Eq.\ (\ref{eq:eps_fadin_kuraev}) allows us to rewrite the
Eq.\ (\ref{eq:fadin_kuraev_formula_0}) in terms of the detection efficiency
at each c.m.\ energy:
\begin{align}
  \sigma_{\rm vis}(s) = \varepsilon(s)\int\limits^{x_0}_0 dx\;
  \sigma_{\rm B}(s(1-x))F(x, s). \label{eq:fadin_kuraev_formula}
\end{align}

The Born cross section at each c.m.\ energy in data has been found by
solving this integral equation. For this goal, the unknown Born cross
section has been interpolated with first-order polynomials from one
c.m.\ energy point to the next one, so the coefficients of the
interpolation polynomials linearly depend on the Born cross section at
each c.m.\ energy. Since the integral in 
Eq.\ (\ref{eq:fadin_kuraev_formula}) can be calculated at each
c.m.\ energy after the interpolation procedure, we can rewrite
Eq.\ (\ref{eq:fadin_kuraev_formula}) as follows:
\begin{align}
  \vec{\sigma}_{\rm vis} = \mathcal{A}\vec{\sigma}_{\rm
    B}, \label{eq:matrix_fk}\\ \vec{\sigma}_{\rm B} =
  \mathcal{A}^{-1}\vec{\sigma}_{\rm vis}, \nonumber
\end{align}
where $\vec{\sigma}_{\rm vis} = (\sigma_{\rm vis}(s_1), \sigma_{\rm
  vis}(s_2), ..., \sigma_{\rm vis}(s_n))$ is the vector composed of
visible cross sections at each c.m.\ energy, $\mathcal{A}$ is the
matrix of the integral operator from
Eq.\ (\ref{eq:fadin_kuraev_formula}), and $\vec{\sigma}_{\rm B} =
(\sigma_{\rm B}(s_1), \sigma_{\rm B}(s_2), ..., \sigma_{\rm B}(s_n))$
is the vector of numerical solutions for Born cross sections at each
c.m.\ energy. The first c.m.\ energy point used in the cross section 
interpolation
equals the $\SP$ threshold ($\sqrt{s} = 2m_{\pi} + m_{\eta}$). The Born cross
section and its uncertainty at this point are equal to zero. The
inverse error matrix~\cite{inv-err-matrix} for the Born cross section can
be calculated using the following formula:
\begin{align}
  \mathcal{M} = \mathcal{A}^{T}\Lambda \mathcal{A},
  \label{eq:born-cs-error-matrix}
\end{align}
where $\Lambda$ is a diagonal inverse error matrix for the visible cross
section. The c.m. energy, $\SP$ Born cross section, $\SP$ yield,
detection efficiency and integrated luminosity are listed in
Table~\ref{tab-result}.
\begin{figure}
  \centering \includegraphics[width=0.8\linewidth]{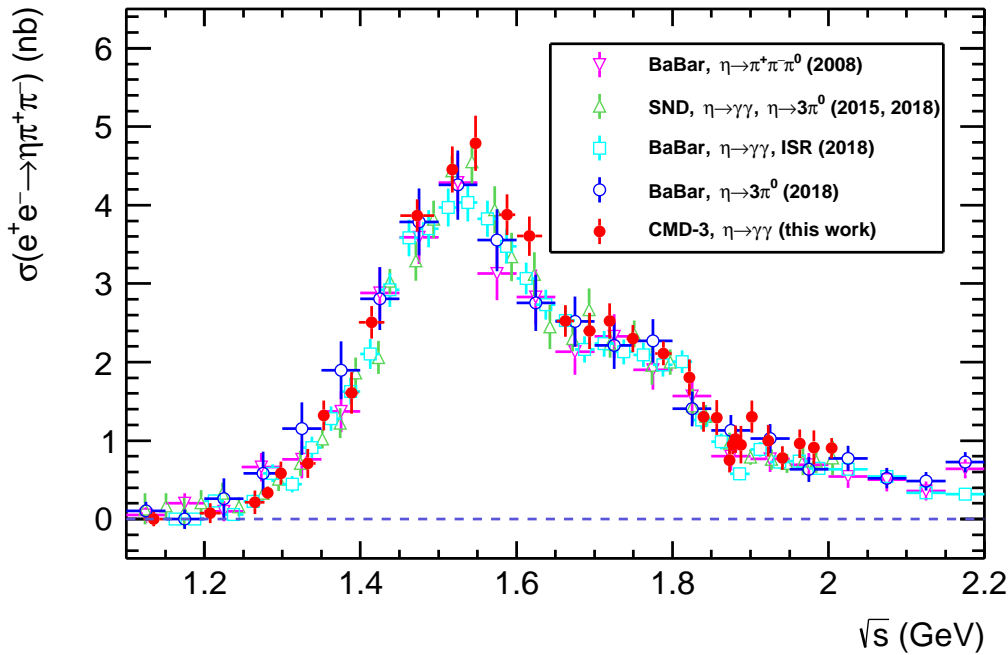}
  \caption{Born cross section for $\SP$ measured at the CMD-3, SND and
    BaBar. The vertical errors for the Born cross section measured at
    the CMD-3 correspond to square roots of the error matrix diagonal
    elements. The close points of the cross section measured with the
    CMD-3 detector are combined together.
    \label{fig:cs}}
\end{figure}
In order to compare the result of the $\SP$ cross section measurement
with the previous measurements, we combine the close c.m.\ energy points in the
cross section measured with the CMD-3. The corresponding energy dependence of
the $\SP$ Born cross section is shown in Fig.\ \ref{fig:cs}. The $\SP$ cross
section values at the combined c.m.~energy points are also listed in
Table~\ref{tab-cs-combined}.

The total systematic uncertainty of the Born cross section is about
$6.0\%$ and consists of the contributions from the following sources:
the detection efficiency ($5.7\%$), the uncertainty of the ISR
correction~\cite{kuraev-fadin} ($0.1\%$), the uncertainty related with
the FSR influence on the detection efficiency ($0.5\%$), the uncertainty on the
integrated luminosity ($1\%$), and the uncertainty of the Born cross section
numerical calculation ($1\%$). The systematic uncertainty on the detection
efficiency includes the following contributions:
\begin{itemize}
\item trigger efficiency,
\item the requirement on $\chi^2$ of the kinematic fit,
\item charged pion reconstruction efficiency,
\item photon reconstruction efficiency,
\item use of the $\SP$ cross section measured with BaBar for MC simulation 
of the studied process.
\end{itemize}

The trigger efficiency uncertainty ($0.1$--$0.9\%$) has been estimated
as the error of the fit assuming a constant function for the energy
dependence of the trigger efficiency correction, $\delta_{\rm trigg}$.

The uncertainty related to the requirement on $\chi^2$ of the kinematic fit
($1.1\%$) has been estimated as the error of $\delta_{\chi^2}$
obtained using Eq.\ (\ref{eq:chi2-correction}) and two statistically
independent $\chi^2$ regions, $\chi^2<30$ and $30<\chi^2<50$.

The uncertainty of the reconstruction efficiency for charged pions
($5.6\%$) has been estimated as the maximum uncertainty for all
c.m.\ energy points given by the uncertainty propagation formula, applied
to Eq.\ (\ref{eq:track-correction}).

The uncertainty of the reconstruction efficiency for photons ($0.2\%$)
has been estimated as the maximum uncertainty for all c.m.\ energy
points given by the uncertainty propagation formula, applied to
Eq.\ (\ref{eq:photon-correction}).

The uncertainty due to use of the $\SP$ cross section measured with 
BaBar to simulate ISR has been estimated as the relative difference of
detection efficiencies in cases of using $\SP$ cross sections measured
with BaBar and CMD-3 in MC simulation. The value of this
uncertainty ($0.4\%$) appears to be less than its statistical error
($1.2\%$) and is neglected.

The uncertainty related to the shape of the background distribution in
two-photon invariant mass spectra has been estimated as the relative
difference between the $e^+e^-\ra\eta\pipi$ yields, $(N_2 - N_1) / N_1
= (0.5\pm1.4)\%$, found from the fit to two-photon invariant mass
spectra using two different background distribution functions.  The
first function is a first-order polynomial, the second one is the
background distribution taken from multihadron MC
simulation~\cite{cmd3-multihadron}. The difference between the
$e^+e^-\ra\eta\pipi$ yields corresponding to these background hypotheses
is found to be statistically insignificant and neglected.

The uncertainty related with the FSR influence on the detection efficiency
has been estimated using the PHOTOS++ package~\cite{photospp-2010, photos-2007}.
To obtain this uncertainty, the detection efficiencies for two kinds of the
$\SP$ MC simulations have been compared at several c.m. energy points. The first
kind of $\SP$ MC simulation does not take FSR into account and is described in
Sec.~\ref{sec:mc-simulation}. The second kind of the $\SP$ MC simulation is
the same as the first one, but it also takes FSR into account. It has been
found that the upper limit on the corresponding uncertainty is $0.5\%$.

The uncertainty of the Born cross section numerical calculation has
been estimated using the following formula:
\begin{align}
  \sigma_{\rm calc} = |(\mathcal{A}^{-1}\sigma^{\rm fit}_{\rm vis} -
  \sigma^{\rm VMD}_{\rm B}) / \sigma^{\rm VMD}_{\rm B}|,
  \label{eq:calc-error}
\end{align}
where the matrix $\mathcal{A}$ has been taken from
Eq.\ (\ref{eq:matrix_fk}), $\sigma^{\rm fit}_{\rm vis}$ is the fit of the
visible cross section in the vector meson dominance model (VMD),
$\sigma^{\rm VMD}_{\rm B}$ is the VMD parametrization of the Born
cross section obtained form the fit of the visible cross section. The
visible cross section has been fitted using
Eq.\ (\ref{eq:fadin_kuraev_formula}) and VDM parametrization of the Born
cross section in three different ways, discussed below. The obtained
uncertainty depends on c.m.\ energy in the following way:
\begin{align}
  \sigma_{\rm calc} =
  \begin{cases}
    1.0\%,\; \sqrt{s}\le1.35\;{\rm GeV}\\ 0.2\%,\;
    1.35\;{\rm GeV}<\sqrt{s}\le2.01\;{\rm GeV}
  \end{cases},
\end{align}
where a relatively big uncertainty at c.m.\ energies $\sqrt{s}\le1.35$
GeV is due to the unknown threshold behavior of the cross section.

The sources of the systematic uncertainty and their contributions are
listed in Table~\ref{tab-sys}.
\begin{table}[t]
  \caption{The sources of the systematic uncertainty.}
  \label{tab-sys}
  \begin{ruledtabular}
    \begin{tabular}[t]{lcc}
      Source & \multicolumn{2}{c}{Uncertainty, $\%$}\\ &
      $\sqrt{s}\leq1.35$ GeV & $\sqrt{s} > 1.35$ GeV\\ [0.5ex] \hline
      \\ [-1.5ex] $\chi^2$ selection criterion & \multicolumn{2}{c}{$1.1$} \\
      Reconstruction of charged pions & \multicolumn{2}{c}{$5.6$}\\
      Photon reconstruction & \multicolumn{2}{c}{$0.2$}\\
      Luminosity & \multicolumn{2}{c}{$1.0$}\\
      ISR correction & \multicolumn{2}{c}{$0.1$}\\
      FSR & \multicolumn{2}{c}{$0.5$}\\
      Trigger efficiency & $0.9$ & $0.1$\\
      Uncertainty of the Born cross section numerical calculation & 1.0 & 0.2\\
      [1.0ex] \hline Total uncertainty & \multicolumn{2}{c}{$6.0$}
    \end{tabular}
  \end{ruledtabular}
\end{table}

The function used for the parametrization of the $\SP$ Born cross
section in the VMD model contains contributions of several isovector
resonances $\ro$, $\rop$, $\ropp$ that decay to the $\eta\ro$ final
state~\cite{achasov-karnakov-1984, achasov-kozhevnikov-1997} (an isoscalar one is suppressed by G-parity conservation):
\begin{align}
  \label{eq:born-cross-section-VMD}
  \sigma_{B}(s) =
  \frac{4\alpha^{2}}{3s\sqrt{s}}\mathcal{I}(s)|\mathcal{F}(s)|^{2},\\ \mathcal{I}(s)
  = \displaystyle\int\limits_{4m_{\pi}^{2}}^{\big(\sqrt{s} -
    m_{\eta}\big)^{2}}dq^2\frac{\sqrt{q^2}
    \Gamma_{\ro}(q^2)P_{\eta}^{3}(s,q^{2})}{\big(q^2-m_{\ro}^{2}\big)^2
    + \big(\sqrt{q^2}\Gamma_{\ro}(q^2)\big)^2}, \nonumber\\ P_{\eta}^2
  = \frac{\big(s-m_{\eta}^2-q^2\big)^2 - 4m_{\eta}^2q^2}{4s},\nonumber
\end{align}
where $m_{\ro}$ is the $\ro$ mass, $\Gamma_{\ro}(q^2)$ is the energy-dependent
$\ro$ width, $q^2$ is the square of the $\pipi$
invariant mass and the form factor $\mathcal{F}(s)$ corresponds to the
transition $\gamma^{*}\ra\eta\ro$:
\begin{align}
  \mathcal{F}(s) = \sum_{V} \frac{m_{V}^2}{g_{V\gamma}}
  \frac{g_{V\rho\eta}}{s-m_{V}^2+i\sqrt{s}\Gamma_{V}(s)}, \label{eq:form-factor}\\ V
  = \ro,\;\rop,\;\ropp \nonumber.
\end{align}
The following formula describes the energy dependence of the
$\ro$ width:
\begin{align}
  \Gamma_{\ro}(q^2) = \Gamma_{\ro}(m_{\ro}^2)\frac{m_{\ro}^2}{q^2}
  \Big(\frac{p^2_{\pi}(q^2)}{p^2_{\pi}(m_{\ro}^2)}\Big)^{\frac{3}{2}},
\end{align}
where $p^2_{\pi}(q^2)$ is the momentum of each pion from $\ro\ra\pipi$:
\begin{align}
  p^2_{\pi}(q^2) = q^2/4 - m_{\pi}^2.
\end{align}

The following formula is used to describe energy dependencies of the $\rop$
and $\ropp$ widths:
\begin{align}
  \Gamma_{\rm V^{\prime}}(s) = \Gamma_{{\rm V^{\prime}}\ra\pipi}(s)
  C^2_{\rm VPP}(s) + \Gamma_{{\rm V^{\prime}}\ra\omega\pi^0}(s)
  C^2_{\rm VVP}(s) + \Gamma_{{\rm V^{\prime}}\ra4\pi}(s) C^2_{\rm
    4\pi}(s),
\end{align}
where $\rm V^{\prime}$ is $\rop$ or $\ropp$, $\Gamma_{{\rm
    V^{\prime}}\ra\pipi}(s)$ is the energy-dependent ${\rm
  V^{\prime}}\ra\pipi$ decay width, $\Gamma_{{\rm
    V^{\prime}}\ra\omega\pi^0}(s)$ is the energy-dependent 
${\rm V^{\prime}}\ra\omega\pi^0$ decay width and $\Gamma_{{\rm
    V^{\prime}}\ra4\pi}$ is the energy-dependent ${\rm
  V^{\prime}}\ra4\pi$ decay width. The energy dependence of the ${\rm
  V^{\prime}}\ra\pipi$ decay width has been described using the
following formula:
\begin{align}
  \Gamma_{{\rm V^{\prime}}\ra\pipi}(s) = \mathcal{B}({\rm
    V^{\prime}}\ra\pipi)\Gamma_{\rm V^{\prime}}(m_{\rm
    V^{\prime}}^2)\frac{m_{\rm V^{\prime}}^2}{s}
  \Big(\frac{p^2_{\pi}(s)}{p^2_{\pi}(m_{\rm
      V^{\prime}}^2)}\Big)^{\frac{3}{2}},
\end{align}
where $\mathcal{B}({\rm V^{\prime}}\ra\pipi)$ is the branching
fraction of the ${\rm V^{\prime}}\ra\pipi$ decay. The energy
dependence of the ${\rm V^{\prime}}\ra\omega\pi^0$ can be written in
the following form:
\begin{align}
  \Gamma_{{\rm V^{\prime}}\ra\omega\pi^0}(s) = \mathcal{B}({\rm
    V^{\prime}}\ra\omega\pi^0)\Gamma_{\rm V^{\prime}}
  \Big(\frac{p^2_{\omega}(s)}{p^2_{\omega}(m_{\rm
      V^{\prime}}^2)}\Big)^{\frac{3}{2}},
\end{align}
where $p_{\omega}$ is the momentum of each particle from the final state
of ${\rm V^{\prime}}\ra\omega\pi^0$ decay in the c.m.\ frame:
\begin{align}
  p^2_{\omega} (s) = (s - (m_{\omega} + m_{\pi})^2)(s - (m_{\omega} -
  m_{\pi})^2) / (4s).
\end{align}
The energy dependence of the ${\rm V^{\prime}}\ra4\pi$ decay width can be 
estimated using phase space:
\begin{align}
  \Gamma_{{\rm V^{\prime}}\ra4\pi} = \mathcal{B}({\rm
    V^{\prime}}\ra4\pi)\Gamma_{\rm
    V^{\prime}}\frac{\Phi_{4\pi}(s)}{\Phi_{4\pi}(m^2_{\rm
      V^{\prime}})}\sqrt{\frac{m^2_{\rm V^{\prime}}}{s}},
\end{align}
where $\Phi_{4\pi}$ is the phase space of $4\pi$.  The functions
$C_{\rm VPP}(s)$, $C_{\rm VVP}(s)$ and $C_{4\pi}(s)$ are the
corresponding Blatt-Weisskopf barrier factors:
\begin{align}
  C^2_{\rm VPP}(s) = \frac{1 + r^2_0p^2_{\pi}(m^2_{\rm V^{\prime}})}{1
  + r^2_0p^2_{\pi}(s)},\\
  C^2_{\rm VVP}(s) = \frac{1 +
    r^2_0p^2_{\omega}(m^2_{\rm V^{\prime}})}{1 +
  r^2_0p^2_{\omega}(s)},\nonumber\\
  C^2_{4\pi}(s) = \frac {1 +
  r^2_0(m^2_{V^{\prime}}-(4m_{\pi})^2)/4} {1 + r^2_0(s -
  (4m_{\pi})^2)/4},\nonumber
\end{align}
where the effective interaction radius, $r_0$, has been taken equal to
$2.5$ GeV$^{-1}$. Typical values of $r_0$ used in other papers are
$2$--$4$ GeV$^{-1}$~\cite{jamin, snd-3pi-2003}.

\begin{table}[t]
  \caption{Table of parameters extracted from the fit of the $\SP$
    cross section in the VMD model.  Parameters listed without
    uncertainties are fixed. All listed uncertainties are
    statistical.}
  \label{tab-csfit}
  \begin{ruledtabular}
    \begin{tabular}[t]{lcccc}
      Parameters & Model 1, solution 1 & Model 1, solution 2 & Model 2, solution
                                                               1 & Model 2, solution 2\\ [0.5ex]
      \hline \\ [-1.5ex]
      \DTLforeach{cross-section-fit}{\pars=pars,\units=units,\ma=m1,\mb=m2,\mc=m3,\md=m4}%
                 {%
                   \ifthenelse{\DTLiseq{\units}{---}}{$\pars$}{$\pars$,
                     \units} &
                   \ifthenelse{\DTLiseq{\ma}{---}}{\ma}{$\ma$} &
                   \ifthenelse{\DTLiseq{\mb}{---}}{\mb}{$\mb$} &
                   \ifthenelse{\DTLiseq{\mc}{---}}{\mc}{$\mc$} &
                   \ifthenelse{\DTLiseq{\md}{---}}{\md}{$\md$}
                   \DTLiflastrow{}{\\} }\\ [1.0ex]
    \end{tabular}
  \end{ruledtabular}
\end{table}

According to Ref.\ \cite{pdg}, the following relations hold between
the different $\rop$ and $\ropp$ decay modes:
\begin{align}
  \frac{\Gamma(\rop\ra\pi\pi)}{\Gamma(\rop\ra{4\pi})} = 0.37\pm0.1,\\
  \frac{\Gamma(\rop\ra\pi\pi)}{\Gamma(\rop\ra\omega\pi)}\sim0.32,\nonumber\\
  \frac{\Gamma(\ropp\ra\pi\pi)}{\Gamma(\ropp\ra{4\pi})} = 0.16\pm0.04.\nonumber
\end{align}
Assuming that $\mathcal{B}(V^{\prime}\ra\pipi) + \mathcal{B}(V^{\prime}\ra\omega\pi^0) +
\mathcal{B}(V^{\prime}\ra{4\pi}) = 1$ and taking into account that the decay
$\ropp\ra\omega\pi$ is not seen~\cite{pdg}, we estimate $\mathcal{B}(V^{\prime}\ra\pipi)$,
$\mathcal{B}(V^{\prime}\ra\omega\pi^0)$ and $\mathcal{B}(V^{\prime}\ra4\pi)$
branching fractions as follows:
\begin{align}
  \mathcal{B}(\rop\ra\pipi) = 15\%,\\
  \mathcal{B}(\rop\ra\omega\pi^0) = 45\%,\nonumber\\
  \mathcal{B}(\rop\ra{4\pi}) = 40\%,\nonumber\\
  \mathcal{B}(\ropp\ra\pipi) = 14\%,\nonumber\\
  \mathcal{B}(\ropp\ra\omega\pi^0) = 0\%,\nonumber\\
  \mathcal{B}(\ropp\ra{4\pi}) = 86\%.\nonumber
\end{align}
While fitting the $\SP$ Born
cross section, the branching fractions of the $\rop$ and
$\ropp$ are fixed at these values.

The parameters $g_{V\rho\eta}$ and $g_{V\gamma}$ are the coupling
constants for the transitions $V\ra\rho\eta$ and $V\ra\gamma^{*}$ and
can be redefined as $g_{V\rho\eta}/g_{V\gamma} = g_{V} e^{i\phi_{V}}$.
The value of the constant $g_{\ro}$ related to $\ro\ra\ro\eta$ is
calculated using data on the partial width for the decay
$\ro\ra\eta\gamma$~\cite{pdg}:
\begin{align}
  \label{eq:couplings}
  g^2_{\ro} = \frac{24}{\alpha}m^3_{\ro}\frac{\Gamma(\rho\ra\eta\gamma)}{\big(m^2_{\ro}-m^2_\eta\big)^3},\\ g_{\ro}\approx1.586\;\rm
  GeV^{-1}.\nonumber
\end{align}

Mass and width of the $\ro$ resonance are fixed at their nominal
values~\cite{pdg}.  Masses and widths of other resonances are allowed
to vary within their errors. The phase of the $\ro$ is taken to be $0$.

The $\SP$ Born cross section data has been fitted within several modes
using $\chi^2$ minimization:
\begin{align}
  \label{eq:born-chi-square}
  \chi^2_{\sigma_{\rm B}} = (\vec{\sigma}_{\rm B} -
  \vec{f})^T\mathcal{M}(\vec{\sigma}_{\rm B} - \vec{f}),
\end{align}
where $\mathcal{M}$ is the error matrix for the Born cross section
(Eq.\ (\ref{eq:born-cs-error-matrix})), $\vec{f}= (f(s_1), f(s_2), ...,
f(s_n))$ is the vector of values for the function describing the Born
cross section within a certain
  model. We consider two models. One of them contains contributions
of the $\ro$ and $\rop$ resonances to the transition form factor
$\mathcal{F}(s)$ while another one contains also the contribution of the 
$\ropp$. Further, these models will be referred as ``Model 1'' and ``Model 2'',
respectively.

One also has to take into account a well-known fact about the ambiguity 
of determination of parameters for a few
interfering resonances. According to Ref.~\cite{fit-ambiguity}, 
$2^{n - 1}$ local minima for the fit to the cross section are expected, 
where $n$ is the number 
of resonances. This formula has been obtained under the assumption that the 
widths of the resonances do
not depend on energy. In this work, two local minima were actually obtained 
for the fit in the case of the $\ro$ and the $\rop$ presence. Further, 
these local minima are referred to as ``Model 1, solution 1'' and ``Model 1, solution 2''. 
When the $\ropp$ contribution is also taken into
account, two local minima are observed instead of four. In the following,
they are referred to as ``Model 2, solution 1'' and ``Model 2, solution 2''.
The fact that there are two local minima only is probably due to width energy
dependence and the cross section uncertainties.

\begin{figure}
  \centering \includegraphics[width=0.8\linewidth]{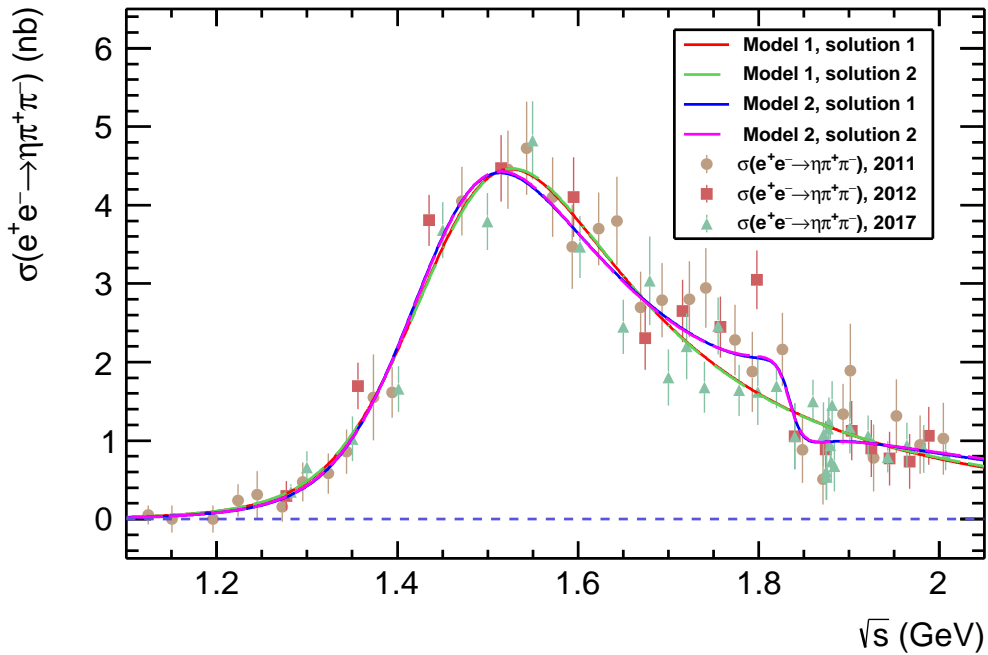}
  \caption{The $\SP$ Born cross section (points with error bars)
    measured with the CMD-3 detector and fitted with ``Model 1, solution 1'' (red
    solid curve), ``Model 1, solution 2'' (blue dashed curve), ``Model 2, solution 1''
    (magenta solid curve) and ``Model 2, solution 2'' (green dashed curve).  The
    ``Model 1, solution 1'' and ``Model 1, solution 2'' include contributions of $\ro$ and
    $\rop$ and correspond to two different local minima of the fit
    chi-square. The``Model 2, solution 1'' and ``Model 2, solution 2'' include contributions
    of $\ro$, $\rop$ and $\ropp$ and correspond to two different local
    minima of the fit chi-square.\label{fig:cs_fit}}
\end{figure}

The results of the fits are presented in Table~\ref{tab-csfit} and
shown in Fig.\ \ref{fig:cs_fit}. The fits within the model, where 
the $\ropp$ contribution is taken into account, have a better quality than
those within the model without the $\ropp$ contribution.

Using parameters $\Gamma(\rop\ra\ee)\mathcal{B}(\rop\ra\epp)$ and
$\Gamma(\ropp\ra\ee)\mathcal{B}(\ropp\ra\epp)$ instead of parameters
$g_{\rop}$ and $g_{\ropp}$ and the relation
\begin{align}
  \label{eq:width-branching}
  \Gamma(V^{\prime}\ra\ee)\mathcal{B}(V^{\prime}\ra\epp) =
  \frac{\alpha^2}{9\pi}
  \frac{|g_{V^{\prime}}|^2m_{V^{\prime}}}{\Gamma_{V^{\prime}}}\mathcal{I}(m^2_{V^{\prime}}),
\end{align}
we perform fits corresponding to solutions
$1$--$2$ in each of the two models. The integral $\mathcal{I}$ has been defined in
Eq.\ (\ref{eq:born-cross-section-VMD}), $V^{\prime}$ is $\rop$ or
$\ropp$, $\Gamma_{V^{\prime}}$ is width of $V^{\prime}$ at
$V^{\prime}$ mass, $m_{V^{\prime}}$.  The fit results for the
$\Gamma(\rop\ra\ee)\mathcal{B}(\rop\ra\epp)$ and
$\Gamma(\ropp\ra\ee)\mathcal{B}(\ropp\ra\epp)$ products are presented
in Table~\ref{table:products1450} and Table~\ref{table:products1700},
respectively.

\begin{table}[t]
  \caption{The $\Gamma(\rop\ra\ee)\mathcal{B}(\rop\ra\epp)$ products
    obtained from different fits, which correspond to solutions
    $1$--$2$ in each of the two models. The first uncertainty in each product
    is statistical, the last one is systematic.}
  \label{table:products1450}
  \begin{ruledtabular}
    \begin{tabular}[t]{lc}
      Solution & $\Gamma(\rop\ra\ee)\mathcal{B}(\rop\ra\epp)$, eV\\ \hline
      \\ [-1.5ex]
      \DTLforeach{products-rho1450}{\solution=solution,\wbproduct=wbproduct}
      { \solution & \wbproduct
                 \DTLiflastrow{\\ [-2.5ex]}{\\} }
      \end{tabular}
  \end{ruledtabular}
\end{table}

\begin{table}[t]
  \caption{The $\Gamma(\ropp\ra\ee)\mathcal{B}(\ropp\ra\epp)$ obtained from
    the different fits, which correspond to solutions $1$--$2$ of the second
    model. The first uncertainty in each product is statistical, the last one is
    systematic.}
  \label{table:products1700}
  \begin{ruledtabular}
    \begin{tabular}[t]{lc}
      Solution &$\Gamma(\ropp\ra\ee)\mathcal{B}(\ropp\ra\epp)$, eV\\ \hline
      \\ [-1.5ex]
      \DTLforeach{products-rho1700}{\solution=solution,\wbproduct=wbproduct}
                 { \solution & \wbproduct
                   \DTLiflastrow{\\ [-2.5ex]}{\\} }
      \end{tabular}
  \end{ruledtabular}
\end{table}

The $\SP$ Born cross section can be used to calculate the
$\tau^-\ra\eta\pi^-\pi^0\nu_{\tau}$ branching fraction. To reach this
goal one has to use the following formula, which has been obtained
under the CVC hypothesis~\cite{gilman}:
\begin{align}
  \label{eq:tau-branching-cvc}
  \frac{\mathcal{B}(\tau^-\ra\eta\pi^-\pi^0\nu_{\tau})}{\mathcal{B}(\tau^-\ra\nu_{\tau}e^-\bar{\nu_e})}=\frac{3\cos^2\theta_C}{2\pi\alpha^2m^8_{\tau}}\int\limits^{m^2_{\tau}}_0
  dq^2 q^2
  \big(m^2_{\tau}-q^2\big)^2\big(m^2_{\tau}+2q^2\big)\sigma_{\SP}(q^2).
\end{align}
The calculation of the $\tau^-\ra\eta\pi^-\pi^0\nu_{\tau}$ branching fraction 
using the CMD-3 data leads to the following result:
\begin{align}
  \label{eq:tau-branching-cmd3}
  \mathcal{B}(\tau^-\ra\eta\pi^-\pi^0\nu_{\tau}) = (0.168\pm0.006\pm0.011)\%,
\end{align}
where the first uncertainty is statistical and the second is systematic.
This result can be compared with the world average value
$\big(0.139\pm0.01\big)\%$~\cite{pdg}, the BaBar result $\big(0.163\pm0.008\big)\%$~\cite{babar-2018-isr},
the SND result $\big(0.156\pm0.004\pm0.010\big)\%$~\cite{snd-2015} and with the CVC
result $\big(0.153\pm0.018\big)\%$ based on the earlier $\SP$
data~\cite{cvc-2011}.

\section{Summary}
The $\SP$ cross section has been measured with the CMD-3 detector in
the c.m.\ energy range $1.2$--$2.0$ GeV using the $\eta$ decay mode
$\eta\ra\gg$.  The obtained result confirms previous $\SP$ cross
section measurements.

The internal structure of the $\eta\pipi$ final state has been studied. 
It has been confirmed that the $\eta\ro$ intermediate state is
dominant.

The fit of the $\SP$ cross section data has been performed within
the two models. One of them includes contributions of the $\ro\ra\ro\eta$
and $\rop\ra\ro\eta$ intermediate mechanisms while the other one includes
also a contribution of the $\ropp\ra\ro\eta$. It has been found that
there are a few local minima of the fit to the cross section depending on
the choice of initial fit parameters.

The products $\Gamma(\rop\ra\ee)\mathcal{B}(\rop\ra\epp)$ and
$\Gamma(\ropp\ra\ee)\mathcal{B}(\ropp\ra\epp)$ corresponding to
each model and fit local minima were also found. The results for these 
products are
listed in Tables~\ref{table:products1450},~\ref{table:products1700}.
The fits to the $\SP$ cross section data have been also used to calculate
the $\tau^-\ra\eta\pi^-\pi^0\nu_{\tau}$ branching fraction under the CVC
hypothesis. The $\tau^-\ra\eta\pi^-\pi^0\nu_{\tau}$ branching fraction
predicted using the $\SP$ cross section data obtained with the CMD-3
detector agrees with the similar SND and BaBar predictions,
and differs by $1.8$ standard deviations of the combined error from the world
average value.

\section{Acknowledgments}
The authors are grateful to the VEPP-2000 team for excellent machine 
operation.
The work has been partially supported by the Russian Foundation for 
Basic Research grant No. 18-32-01020. Part of the work related to
the multihadronic generator is supported by the grant of Ministry of Science
and Higher Education No. 14.W03.31.0026. 

\bibliography{paper-etapipi-bibliography}
\end{document}